\documentclass[draft]{agujournal2019}
\usepackage{url} 
\usepackage{lineno}
\usepackage[inline]{trackchanges} 
\usepackage{soul}
\usepackage{amsmath}
\usepackage{underscore}
\usepackage[utf8]{inputenc}
\usepackage[greek,english]{babel}
\usepackage{CJKutf8}

\journalname{JGR: Space Physics}

\let\oldequation\equation
\let\oldendequation\endequation

\renewenvironment{equation}
  {\linenomathNonumbers\oldequation}
  {\oldendequation\endlinenomath}

\begin{document}
\foreignlanguage{greek}{}\begin{CJK}{UTF8}{gbsn}\end{CJK}

\title{Field Line Curvature (FLC) Scattering in the Dayside Off-Equatorial Minima Regions}


\authors{Bin Cai\affil{1}, Hui Zhu\affil{1}}

\affiliation{1}{Institute of Frontier and Interdisciplinary Science, Shandong University, Qingdao, China}

\correspondingauthor{H. Zhu}{huizhu@email.sdu.edu.cn}




\begin{keypoints}
\item Ring current protons in the dayside off-equatorial regions are significantly affected by FLC scattering.
\item The scattering effects driven by FLC on protons display distinct hemispheric asymmetry in the dayside off-equatorial regions.
\item The dipole tilt angle affects hemispheric asymmetry, thereby influencing the strength and extent of FLC scattering.
\end{keypoints}

%
%

%
%


\begin{abstract}
Magnetic field line curvature (FLC) scattering is an effective mechanism for collisionless particle scattering. In the terrestrial magnetosphere, the FLC scattering plays an essential role in shaping the outer boundary of protons radiation belt, the rapid decay of ring current, and the formation of proton isotropic boundary (IB). However, previous studies have yet to adequately investigate the influence of FLC scattering on charged particles in the Earth's dayside magnetosphere, particularly in the off-equatorial magnetic minima regions. This study employs T89 magnetic field model to investigate the impacts of FLC scattering on ring current protons in the dayside magnetosphere, with a specific focus on the off-equatorial minimum regions. We analyze the spatial distributions of single and dual magnetic minima regions, adiabatic parameter, and pitch angle diffusion coefficients due to FLC scattering as functions of $Kp$. The results show that the effects of FLC scattering are significant not only on the dusk and dawn sides but also in the off-equatorial minima regions on the noon. Additionally, we investigate the role of dipole tilt angle in the hemispheric asymmetry of FLC scattering effects. The dipole tilt angle controls the overall displacement of the dayside magnetosphere, resulting in different FLC scattering effects in the two hemispheres. Our study holds significance for understanding the FLC scattering effects in the off-equatorial region of Earth's dayside magnetosphere and for constructing a more accurate dynamic model of particles.
\end{abstract}


\section{Introduction}

The first adiabatic invariant of a charged particle in a magnetic field is associated with the periodic gyration around the local field line \cite{Northrop1964,Hastie1967,Roederer1970,Lyons1984}. The first adiabatic invariant, characterized by the magnetic moment $\mu = p_{\perp}^2 /(2m_0B)$, is customarily conserved with a certain accuracy. Here $p_{\perp} $ is the particle momentum perpendicular to the magnetic field, $m_0$ is the rest mass, and $B$ is the ambient magnetic field strength. Conservation of $\mu $ mainly depends on the ratio of the time scale of magnetic field variation to charged particle gyration, and the ratio of the spatial scale of the magnetic field inhomogeneity (or the radius of field line curvature $R_c$) to charged particle gyro-radius $\rho$. In a longer time scale ($\gg$ gyro-period), the latter ratio plays a critical role in determining whether the magnetic moment is conserved. The breakdown of the $\mu$ conservation is manifested when the inhomogeneity of magnetic field (the radius of magnetic field line curvature) is comparable to the gyro-radius of charged particles. The physical process discussed in this paper, closely related to the curvature of the field line, leading to the violation of the first adiabatic invariant, is known as $\mu$ scattering or field line curvature (FLC) scattering \cite{Gray1982,Anderson1997,Young2002,Young2008,Birmingham1984}.

A parameter called the adiabatic parameter $\varepsilon = \rho / R_{c}$ has been used to determine the $\mu$ changes induced by FLC scattering, where the quantity is usually calculated at the magnetic equator (i.e., the location of magnetic minimum along field line). When particles reach the magnetic equator, with the weakest magnetic field and the highest perpendicular momentum, the conservation of $\mu$ is easily violated. In general, the condition of the breakdown of $\mu$ conservation is that $\varepsilon $ range from $\sim0.1$ to $\sim1$ \cite{Chirikov1987,Young2008}. For charged particles in the inner magnetosphere, the location of breakdown often occurs on the nightside of the equatorial plane where the magnetic field is weak, and the field line is significantly stretched during geomagnetic storms. In the vicinity of the nightside equator, \citeA{Artemyev2013} investigated the pitch angle diffusion coefficients of electrons around the geostationary orbit ($L \sim 7 $) due to the effects of FLC scattering. They use the current sheet model as the magnetic field, which represents the near-equatorial region at midnight with stretched field lines, and they found the effects of FLC scattering are more important than the effects of wave-particle interactions for high energy ($> $ 1 MeV) electrons. \citeA{Yu2020} considered the mechanism of FLC scattering into a kinetic ring current model and investigated its role in the precipitation of ions into the ionosphere during the 17 March 2013 storm. The simulation results indicate that the process of FLC scattering exerts on energetic and heavy ions on the nightside, and the ions precipitation driven by FLC scattering mainly occurs in the outer region ($L > 4.5 $) on the nightside. Besides, many previous studies related to FLC scattering mainly focus on the nightside of the terrestrial magnetosphere\cite{Gilson2012, Dubyagin2018, Ma2022, Yue2014,Zhu2021}, which is attributed to the occurrence of the magnetic minimum and the most apparent field line stretching in this region. 



The magnetic field characteristics on the dayside magnetosphere are significantly different from those on the nightside. One of the most remarkable features of the dayside magnetosphere is the presence of dual ${B}$ minima along a field line on either side of the equator. Unlike magnetosphere on the nightside have only one B minimum near the equator, the dual ${B}$ minima off the equator lead to the distribution of strength along the field line as a W-shape. When a charged particle drifts into the W-shaped regions, and the strength of the magnetic field at the local maximum near the equator exceeds $B_{m}$, the charged particle becomes temporarily trapped in one of the hemispheres off the equator. Here, the $B_{m}$ denotes the strength of the magnetic field at the particle's magnetic mirror points. This process is called drift orbit bifurcation, and the orbit of particles is known as the Shabansky orbit \cite{Shabansky1968,Shabansky1972,Ozturk2007,Huang2022}. In such magnetic field configuration, the potential FLC scattering may be different from those on the nightside. As far as we know, the investigations on the FLC scattering on the dayside magnetosphere are still lacking.

The remainder of this paper is structured as follows. In Section \ref{method}, we introduce our methodology, including the adopted magnetic field model and the empirical model of pitch angle diffusion coefficients driven by FLC scattering based on \citeA{Young2002}. In Section \ref{result}, we present a series of results related to off-equatorial minima regions, illustrating the spatial distributions and magnitude of FLC scattering effects. Our findings are summarized and discussed in Section \ref{discuss}.

\section{Methodology}
\label{method}
The intensity of FLC scattering, as a diffusion mechanism of charged particles, strongly depends on the configuration and strength of the magnetic field. In this work, we use the dipole and the T89 model \cite{Tsyganenko1989} as the internal and external field to represent the terrestrial magnetic field. By adjusting the $Kp$ index, the T89 field model can be adopted for different geomagnetic conditions, making it suitable for investigating the configuration and magnitude of the dayside magnetosphere. 




To quantitatively assess the impacts of FLC scattering on charged particles, we use the empirical model proposed by \citeA{Young2002} to calculate the pitch angle diffusion coefficient $D_{\alpha \alpha}$ driven by FLC scattering. This model enables a quantitative analysis of the effects of FLC scattering in a realistic magnetosphere, encompassing both active and quiet geomagnetic conditions. The jump of $\mu $ has the general form $\delta\mu = A_0\cos\phi + A_1$ when the particle crosses the equatorial plane every time \cite{Delcourt1994,Anderson1997}, where $\phi $ is the gyrophase of particle and parameters $A_0$ and $A_1$ depend on the field configuration. Following \citeA{Young2002}, $A_0$ is defined as:
\begin{equation}
A_0=e^{c(\varepsilon)}\left(\zeta_1^{a_1(\varepsilon)} \zeta_2^{a_2(\varepsilon)}+C(\varepsilon)\right) \frac{\sin \left(\Lambda(\varepsilon) \alpha\right) \cos ^{\beta(\varepsilon)} \alpha}{\sin \left(\Lambda(\varepsilon) \alpha_{\max }\right) \cos ^{\beta(\varepsilon)} \alpha_{\max }}.
\end{equation}
The symbol $\alpha$ represents the particle's pitch angle at the location of minimum magnetic field strength ($B$ minimum). The coefficients $\zeta_{1}$ and $\zeta_{2}$ are parameters that depend on the specific magnetic field configuration, which is defined as
\begin{equation}
\zeta_{1}=\left.R_{C} \frac{\partial^{2} R_{C}}{\partial S^{2}}\right|_{S_{\max}}, \zeta_{2}=\left.\frac{R_{C}^{2}}{B} \frac{\partial^{2} B}{\partial S^{2}}\right|_{S_{\max}}.
\end{equation}
Here, $S$ represents the arc distance along the field line, and $S_{\max}$ denotes the value of $S$ corresponding to the maximum value of $\varepsilon$. $B(S)$ and $R_{C}(S)$ denote the strength of magnetic field and radius of field curvature as functions of $S$. Coefficients of $C(\varepsilon)$, $c(\varepsilon)$, $a_1(\varepsilon)$, $a_2(\varepsilon)$, $\Lambda(\varepsilon)$, $\beta(\varepsilon)$, and constant $\alpha_{\max }$ are constant polynomials about $\varepsilon$ and can be found in \citeA{Young2002}. By averaging over the ensemble of jumps $\delta \mu$ based on \citeA{Young2008}, the pitch angle diffusion coefficient can be acquired:
\begin{equation}
\label{eq:Young}
 D_{\alpha \alpha}=A_0^2 /\left(T_{b} \sin ^2 \alpha \cos ^2 \alpha\right),
\end{equation}
where the $T_{b} $ denotes the bounce period of a particle. 



The detailed calculation in this work involves the following steps: First, we employ uniform 1D (or 2D) grids in the Cartesian (or polar) coordinate system in a certain direction (or plane), with radial grids ranging from 6 to $12\, R_\text{E}$ (and azimuthal grids ranging from $-0.5\pi$ to $0.5\pi$). Second, we trace the field line from the points (or $Z=0$ plane) to find the location of B-minimum and obtain the corresponding magnetic field intensity, which, in turn, are utilized to derive the adiabatic parameter $ \varepsilon $. If there are dual B minima along a field line, the two sets of $ \varepsilon $ off the equator are recorded. Third, we integrate the particle's parallel velocity along the field line to determine the bounce period $T_{b} $ of a particle with a specific equatorial pitch angle. Considering that the field lines have dual $B$ minima and particles may be trapped in different hemispheres in off-equatorial regions, two equatorial pitch angles are defined, and two sets of $T_b$ are obtained, shown in the next section. Finally, given the parameters of $ \varepsilon $ and $T_b$, the pitch angle diffusion coefficients due to FLC scattering are then calculated based on the \citeA{Young2002} model. Therefore, for off-equatorial regions with dual magnetic minima, two sets of $\varepsilon$, $T_b$, and diffusion coefficients will be computed. In this study, we focus on protons, the dominant component of the ring current, and we choose the energy to be $200$\,keV.

Besides, there is a parameter related to the magnetic field configurations, namely the magnetic dipole tilt angle. The angle between the magnetic axis and the Sun-Earth line is commonly referred to as dipole tilt angle  $\psi=\sin ^{-1}(\hat{\mathbf{s}} \cdot \hat{\mathbf{m}})$. Here, $\hat{\mathbf{m}}$ represents the geomagnetic dipole's northward direction, while $\hat{\mathbf{s}}$ points toward the Sun \cite{laundal2017}. In the GSM coordinates system, it represents the residual angle between the x-axis and the magnetic dipole's north axis. It is conventionally defined as positive when the magnetic dipole's north axis tilts toward the Sun and negative when tilted away. Because the Earth's dipole magnetic axis is inclined at approximately 10.5$^\circ$ from the Earth's rotational axis, the dipole tilt angle shows daily and seasonal variations as the Earth's rotation and revolution. As a benchmark for this study, we start by setting the magnetic dipole tilt angle to -20$^\circ$. In this configuration, the Earth's dayside magnetosphere tilts northward, while the nightside magnetosphere tilts southward. Later, we will discuss the influences of dipole tilt angle on FLC scattering.

\section{Results}
\label{result}
\subsection{FLC scattering on the dayside meridional plane}
\label{result-one-case}

\begin{figure}
\centering
\noindent\includegraphics[width=1\textwidth]{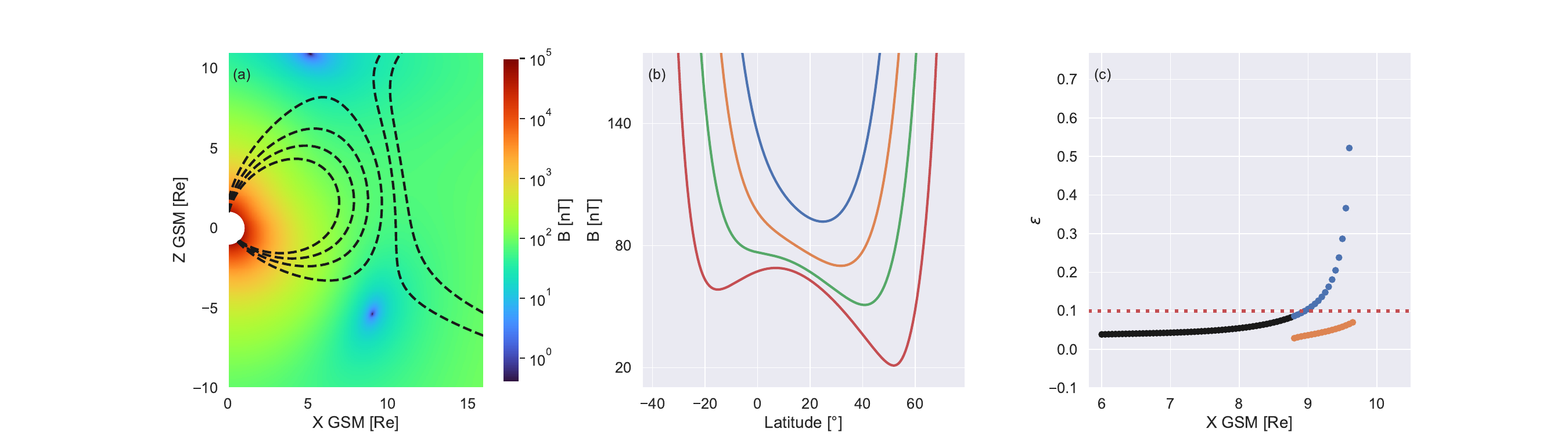}
\caption{
(a) Configuration of the dayside magnetosphere during moderate geomagnetic conditions ($Kp=3$) in the noon-midnight meridional plane. Contours show the magnitude of the magnetic field. The black dashed lines show the field lines starting from X = 6.5, 7.5, 8.5, 9.5, 10.5, and 11.5 $R_\text{E}$. (b) The magnetic field strength as a function of magnetic latitude for four closed field lines (X = 6.5, 7.5, 8.5, 9.5 $R_\text{E}$). (c) A series of $\varepsilon$ are calculated along the X axis between 6 to 10 $R_\text{E}$. The black dots denote regions with a single value of $\varepsilon$. The blue and orange dots denote the value of $\varepsilon$ calculated from the northern and southern hemispheres, respectively. The red dashed horizontal line ($\varepsilon=0.1$) denotes the lower threshold of the FLC scattering that occurs.
}
\label{fig:1}
\end{figure}

Figure \ref{fig:1}(a) shows the configuration of the dayside magnetosphere during moderate geomagnetic conditions ($Kp=3$) in the noon-midnight meridional plane. The black dashed lines show the field lines starting from X = 6.5, 7.5, 8.5, 9.5, 10.5, and 11.5 $R_\text{E}$. Contours show the magnitude of the magnetic field $B$, and two distinct $B$ minima regions off the equator can be seen. The northern one is close to the field line of X = 9.5 $R_\text{E}$. The field lines starting from X = 10.5 and 11.5 correspond to the open field line. Figure \ref{fig:1}(b) shows the magnetic field strength as a function of magnetic latitude for four closed field lines (X = 6.5, 7.5, 8.5, 9.5 $R_\text{E}$). The shape of the distribution of magnetic field strength is W-shape for X = 9.5 $R_\text{E}$ line only, and the others are U-shape. Among the two $B$ minima off the equator, the northern one has a lower value ($\sim$ 20 nT), which is almost half of the south one ($\sim$ 50 nT). A W-shape of the field indicates that the effects of FLC scattering in these regions need to be reconsidered quantitatively.

We can get two kinds of closed field lines by tracing the field line from X = 6 to 10 $R_\text{E}$ at the meridional plane with the step size of $0.05\,R_\text{E}$. One type is characterized by only one $B$ minimum occurring along the line (U-shape). Another type has dual $B$ minima on either side of the equator (W-shape). Correspondingly, the calculations of $\varepsilon $ of proton (200 keV) in this region can also be categorized into two types. It is shown in Figure \ref{fig:1}(c) that the regions between 6 to 8.8 $R_\text{E}$ have single value of $\varepsilon$ by black dots, and the regions between 8.8 to 9.7 $R_\text{E}$ have dual values of $\varepsilon$. Blue and orange dots depict the values of $\varepsilon$ calculated from northern and southern off-equatorial regions. Following the previous studies \cite{Young2008,Yu2020}, we take the $\varepsilon\,=\,0.1$ as the lower threshold of the FLC scattering occurs, indicated by the red dashed horizontal line. In this case, all the values of $\varepsilon $ in the U-shaped regions are below 0.1. In the W-shaped regions, however, the $\varepsilon $ only calculated from the northern hemisphere is greater than 0.1, which increases sharply with distance. The $\varepsilon $ in the southern hemisphere are smaller than 0.1. These results suggest that the effects of FLC scattering may occur only in the northern off-equatorial minima regions, showing an asymmetry between the northern and southern hemispheres. 

\begin{figure}
\centering
\noindent\includegraphics[width=1\textwidth]{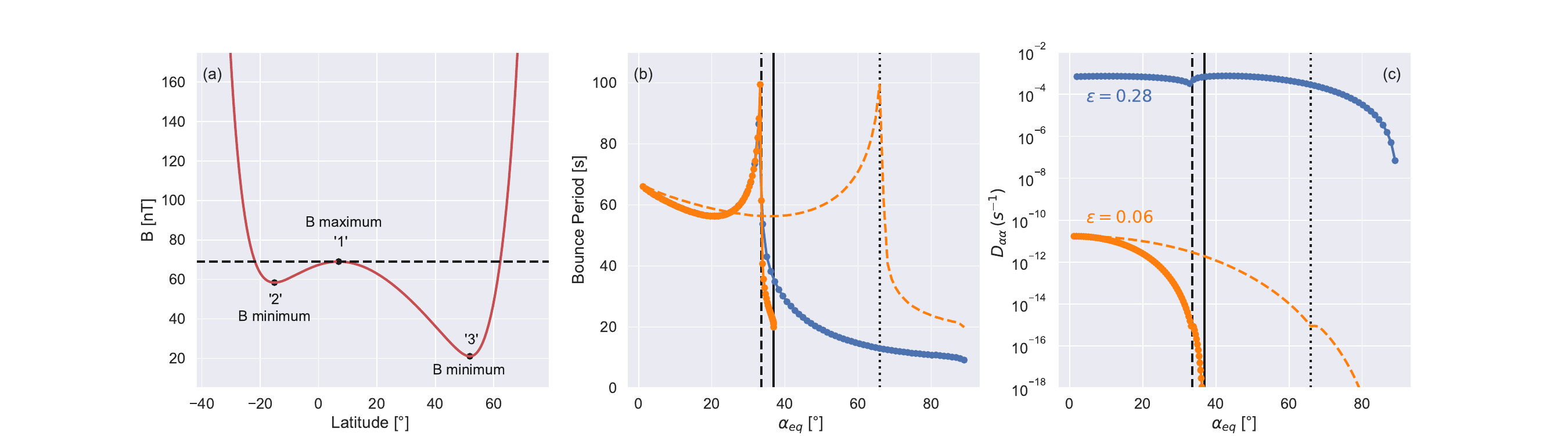}
\caption{(a) Taking field line of X = 9.5 $R_\text{E}$ as an example, the '1', '2', and '3' are labeled to represent the locations of the maximum strength of $B$ between dual minima, the higher and the lower strength of $B$ minima, respectively. (b) The bounce period $T_b$ exhibits variations with equatorial pitch angle. The blue and orange lines (orange dashed line) denote the bounce period of the proton when we assume the location of '3' ('2') as the equator and define the $\alpha_{eq}$ based on it. (c) The pitch angle diffusion coefficients ($D_{\alpha\alpha} $) driven by FLC scattering are shown using the empirical model by \citeA{Young2002}. The blue and orange lines represent the $D_{\alpha\alpha} $ as a function of the equatorial pitch angle defined by the northern and southern $B$ minimum, respectively. Three black vertical lines denote $33.5^\circ$, $36.9^\circ$ and $66.9^\circ$, respectively.}
\label{fig:2}
\end{figure}

We choose the field line starting from X = 9.5 $R_\text{E}$ as an example to illustrate our approach for studying FLC scattering in the off-equatorial minima regions. Figure \ref{fig:2}(a) shows the strength of the magnetic field along the specified field line, same as the red line in Figure \ref{fig:1}(b). Moreover, we label points '1', '2', and '3' to represent the locations of the maximum strength of $B$ between dual minima, the higher and the lower strength of $B$ minima, respectively. The blue dashed line denotes the strength of '1' ($B_1$), which equals to 70 nT. In the \citeA{Young2002} model, the bounce period ($T_b$) is also required to obtain the pitch angle diffusion coefficients. However, the calculations of the bounce period are complicated because of the presence of off-equatorial $B$ minima for the particles trapped off the equator in different hemispheres. Figure \ref{fig:2}(b) illustrates that $T_b$ exhibits a variations with equatorial pitch angle ($\alpha_{eq}$) under two definitions of equator. 




The blue and orange lines denote the bounce period of the proton at $\alpha_{eq}$ from 0 to 90$^{\circ}$ by taking the point '3' as the equator. The particle's bounce motion depends on the strength of magnetic mirror point ($B_m $). When $B_m = B_1$, meaning $\alpha_{eq} = \sin^{-1}\sqrt{B_3/B_1}= 33.5^\circ$, the strength of the magnetic mirror just equals the value of the magnetic maximum. The blue and orange lines reach the peak at this equatorial pitch angle ($33.5^\circ$, marked by the black dashed line). When $B_m > B_1$ ($\alpha_{eq} < 33.5^\circ$), the particle bounces past both points '3' and '2'. When $B_2<B_m < B_1$ ($33.5^\circ< \alpha_{eq} < \sin^{-1}\sqrt{B_3/B_2} = 36.9^\circ$, marked by the black solid line), the particle will be trapped in one of $B$ minimum pockets, bouncing past either point '3' (blue line)  or '2' (orange line). When $B_m < B_2$ ($\alpha_{eq} > 36.9^\circ$), the blue line shows the bounce period of the particle past point '3' only.

However, as the orange dashed line shows, we can also take point '2' as the equator to define the equatorial pitch angle. Similarly, there is a peak of bounce period at $\alpha_{eq} = \sin^{-1}\sqrt{B_2/B_1}= 66.9^\circ$, marked by the black dotted line. When $B_m > B_1$ ($\alpha_{eq} < 66.9^\circ$), the particle bounces past both points '3' and '2'. When $B_m < B_1$ ($\alpha_{eq} > 66.9^\circ$), the orange dashed line shows the bounce period of particle past point '2' only. Overall, unlike the U-shaped field line, the bounce period decreases non-monotonically with a larger pitch angle characterized by a peak at the middle pitch angle. Due to the presence of dual $B$ minima, the particles have two bounce period values when trapped in different hemispheres.

Figure \ref{fig:2}(c) shows the pitch angle diffusion coefficients ($D_{\alpha\alpha} $) driven by FLC scattering using the empirical model by \citeA{Young2002}. The blue and orange lines display the $D_{\alpha\alpha} $ as a function of the equatorial pitch angle defined by taking '3' as the equator. Similarly, the orange dashed line displays the $D_{\alpha\alpha} $ by taking '2' as the equator. The blue and orange lines reflect the influences of FLC scattering on protons in the northern and southern hemispheres, respectively. In the following sections, separate investigations of FLC scattering are performed for the northern and southern hemispheres. Therefore, the pitch angle mentioned below is defined by the corresponding equator of the hemisphere.

By comparing the blue and orange dashed lines, the FLC scattering rates in the northern hemisphere are much stronger than those in the southern hemisphere. For example, at $\alpha_{eq}=30^{\circ} $, the pitch angle diffusion coefficients of FLC scattering are $\sim 4\times10^{-4}\,s^{-1}$ in northern hemisphere and $\sim 4\times10^{-12}\,s^{-1}$ in southern hemisphere, respectively. The difference in the asymmetry of the north-south hemisphere rises from $\varepsilon$. The value of $\varepsilon$ as 0.06 in the southern hemisphere is much smaller than $\varepsilon$=0.28 in the northern. Not only does the difference in the strength of $B$ minimum ($\sim$20 nT and $\sim$60 nT for the north and south, respectively) contribute, but also the difference in the curvature radius of the magnetic field lines at $B$ minimum ($\sim$1.68 $R_E$ and $\sim$2.96 $R_E$ for the north and south, respectively) plays an essential role in determining the different values of $\varepsilon$. This difference, in turn, causes the order of magnitude difference in the $D_{\alpha\alpha}$ values driven by FLC scattering between the northern and southern hemispheres. Overall, we show the magnetic field configurations in off-equatorial minima regions on the meridional plane. By calculating the adiabatic parameter $\varepsilon$ and bounce period $T_b$, we can obtain the pitch angle diffusion coefficients due to FLC scattering. Next, we further investigate the global features of FLC scattering on the dayside magnetosphere.

\subsection{Distribution of $N_{min}$, $\varepsilon$ and diffusion coefficients due to FLC scattering}



\begin{figure}
\centering
\noindent\includegraphics[width=\textwidth]{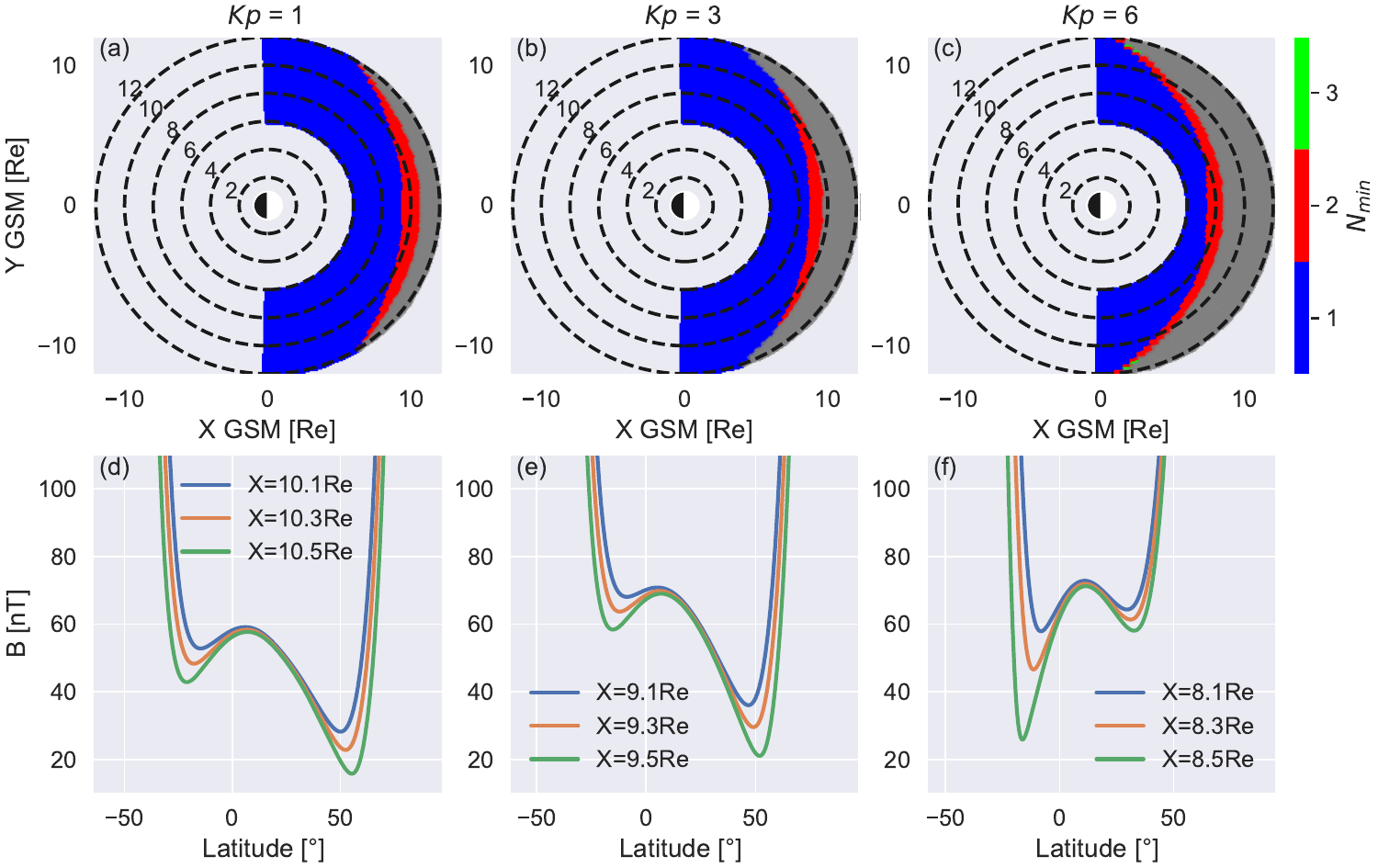}
\caption{(a)-(c) Distribution of the number of $B$ minimum ($N_{min}$) on the $Z=0$ plane under geomagnetic indices $Kp$=1, 3, and 6, respectively.
(d)-(f) The strength of the magnetic field as a function of latitude in the dual $B$ minima regions (the red area in the above rows) corresponding to the $Kp$ = 1, 3, and 6, respectively, and three field lines at the meridional plane with a step size of 0.2 $R_\text{E}$.
}
\label{fig:3}
\end{figure}

We aim to derive the 2D distribution of diffusion coefficients driven by FLC scattering on the $Z=0$ plane of GSM coordinates. Firstly, we need to identify the regions where the configurations of the field line are U-shaped or W-shaped. We employ uniform 2D grids in the polar coordinate system, with 60 radial grids ranging from $6$ to $12\, R_\text{E}$ and 60 azimuthal grids for $3<MLT<18$. Figures \ref{fig:3}(a)-(c) illustrate the distribution of the number of $B$ minimum ($N_{min}$) in T89 field for different geomagnetic indices ($Kp$ = 1, 3, and 6). The regions of $N_{min}=1$ (blue) and 2 (red) correspond to the traced field line characterized by a single $B$ minimum and dual $B$ minima. The gray represents the regions where the field lines are open. The regions having dual $B$ minima off the equator are located between the single $B$ minimum and the open field, and their radial distribution is broader near the noon ($\sim $ 0.5 $R_\text{E}$) than the dusk or down. With the increase of the $Kp$ index, more pronounced compression of the dayside magnetosphere becomes evident. Consequently, this results in the inward movement of regions characterized by closed field lines. The distribution of $N_{min}=2$ has a wide range in the MLT direction. Besides, the field line with three $B$ minima (a third minimum of $B$ occurs near the equator) is found sporadically at $Kp=6$, which is also reported by \citeA{Huang2022}. Because the minimum near the equator is strongly insignificant (not shown here), we consider the three $B$ minima as the dual $B$ minima by considering the off-equatorial minima only.
 
Figures \ref{fig:3}(d)-(f) show the strength of the magnetic field along the field lines in the dual $B$ minima regions (the red area in the above rows) at $Kp$ = 1, 3, and 6, respectively, and three field lines at the meridional plane with a step size of 0.2 $R_\text{E}$ are shown in each panel. From Figures \ref{fig:3}(d) and \ref{fig:3}(e), we can find a similar variation of the $B$ strength with latitude. That is, in the dual magnetic minima, the strength of $B$ is lower in the northern hemisphere. However, at $Kp = 6$, the strength of $B$ is lower in the southern hemisphere. By examining the magnetic field lines, we find that this configuration arises from the fact that the field lines we trace, which are close to the north $B$ minimum, became open. Overall, we obtain the $N_{min}$ distribution on the Z=0 plane in the T89 model and illustrate the locations of dual $B$ minima in off-equatorial regions for different $Kp$ indices.

\begin{figure}
\centering
\noindent\includegraphics[width=\textwidth]{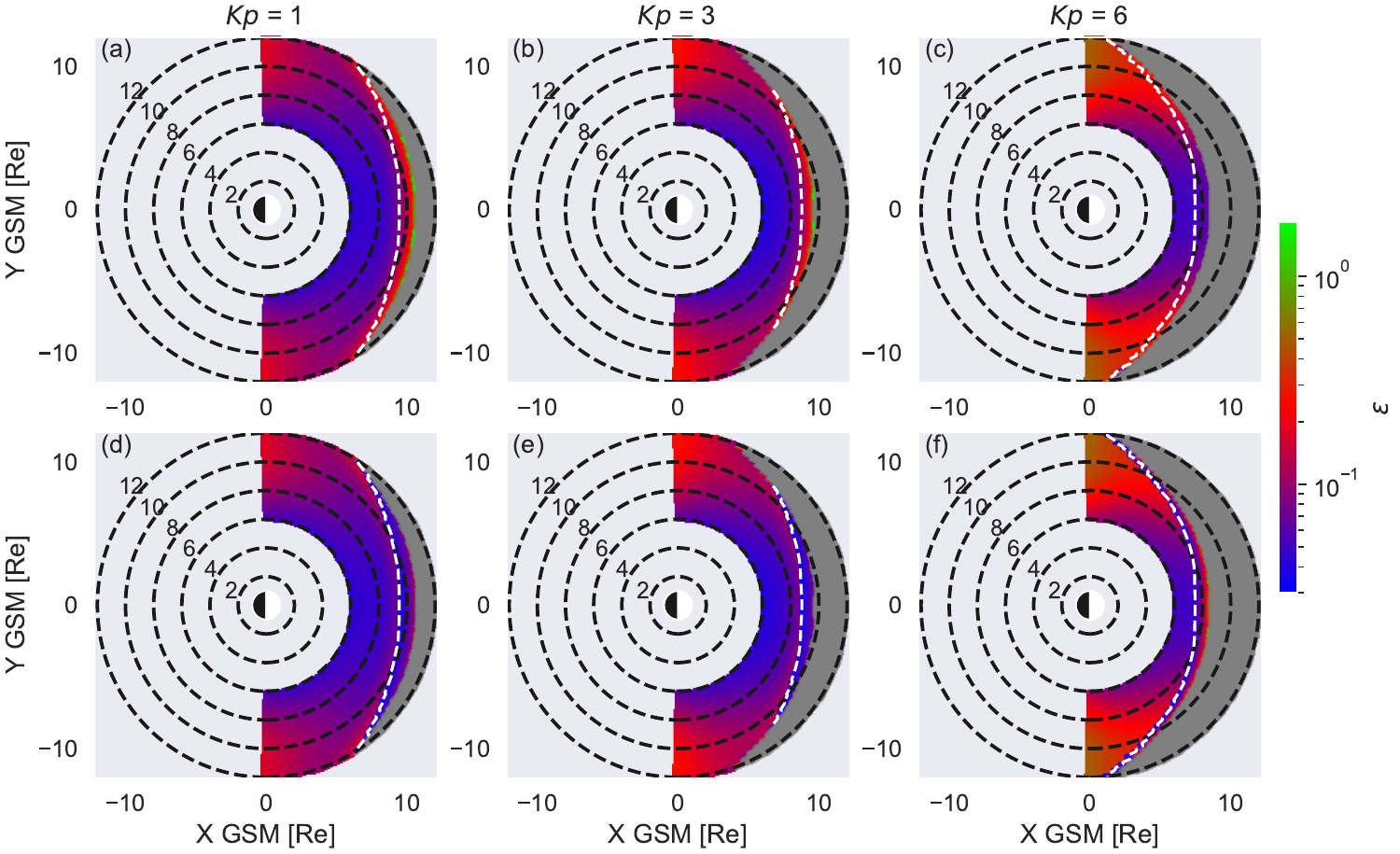}
\caption{Distribution of adiabatic parameter ($\varepsilon$) on the Z=0 plane. The first and second rows represent the values of $\varepsilon$ in the $N_{min}=1$ combined with $N_{min}>1$ regions for the northern and the southern hemispheres, respectively. The white dashed line indicates the boundary between the regions of $N_{min}=1$ and 2.}
\label{fig:4}
\end{figure}

As the critical parameter in determining the influences of FLC scattering, the distribution of $\varepsilon$ on the Z=0 plane is shown in Figure \ref{fig:4}. The white dashed lines represent the boundary between the regions of $N_{min}=1$ and 2 in the Figures \ref{fig:3}(a)-(c). By dividing the off-equatorial regions into northern and southern hemispheres for investigation, the first and second rows represent the values of $\varepsilon$ in the $N_{min}=1$ combined with $N_{min}>1$ regions for the northern and the southern hemisphere, respectively. With the increase of the $Kp$ index, the $\varepsilon$ in all MLT directions becomes larger, concerning the lower magnitude of $B$ and smaller radius of field line curvature. At $Kp=1$ and 3, the regions of $\varepsilon >0.1$ are mainly distributed in the dusk and dawn sections. In addition, in the northern off-equatorial regions, the value of $\varepsilon$ is larger than 0.1 at noon, even reaching the value of 1. For example, comparing Figures \ref{fig:3}(b) and \ref{fig:3}(e), a distinct red region appears outside the white dashed line in the northern hemisphere. However, this tendency is reversed at $Kp=6$, in which southern off-equatorial regions have larger $\varepsilon$ than the northern. This result agrees with the variations of $B$ strength with latitude as shown in Figure \ref{fig:3}(f), in which the smaller magnitude of $B$ is located in the south hemisphere.

\begin{figure}
\centering
\noindent\includegraphics[width=\textwidth]{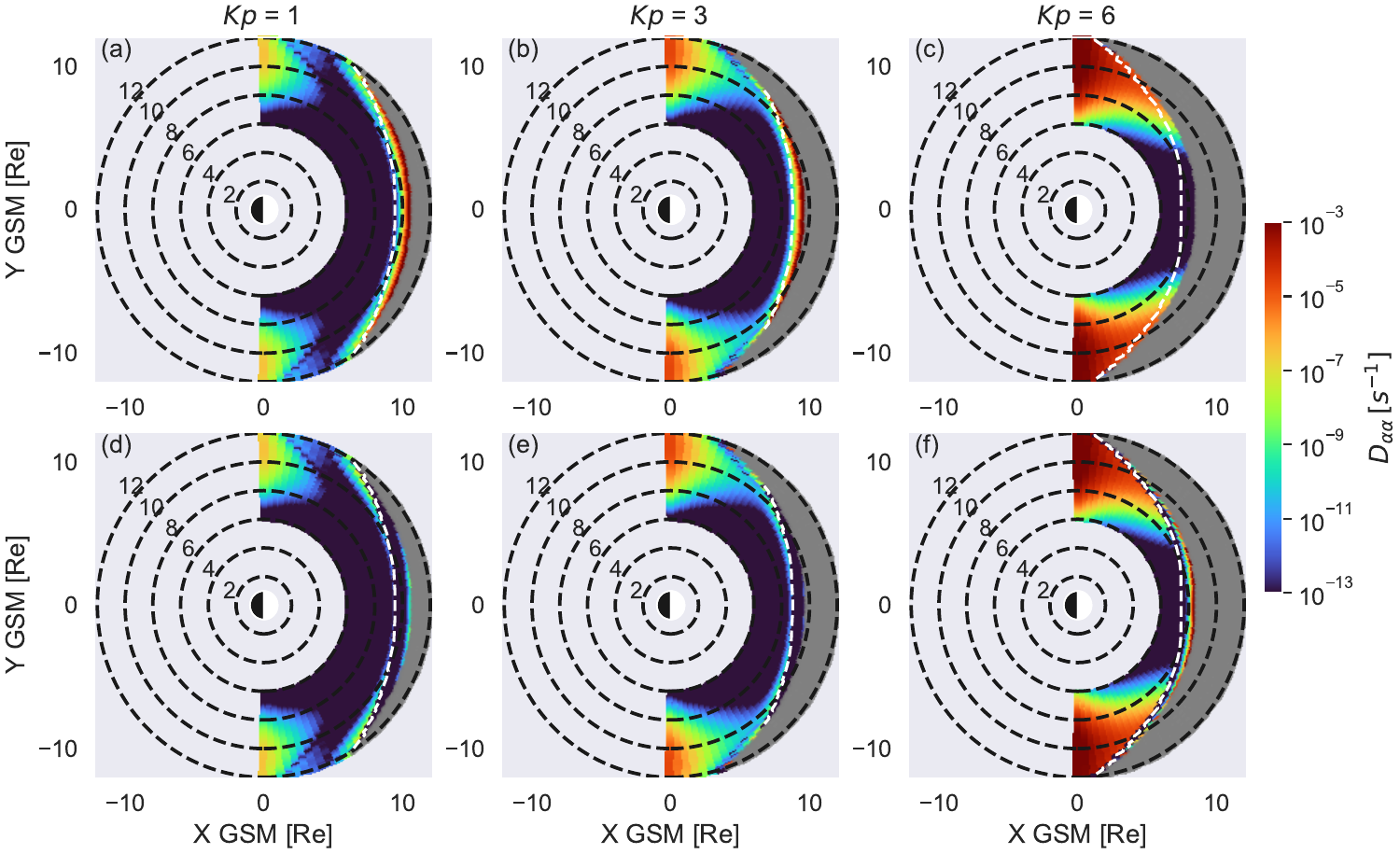}
\caption{The distribution of pitch angle diffusion coefficients ($D_{\alpha\alpha}$) driven by FLC scattering at Z=0 plane using the \citeA{Young2002} model. The first and second rows represent the values of $\varepsilon$ in the $N_{min}=1$ combined with $N_{min}>1$ regions for the northern and the southern hemispheres, respectively. The white dashed line indicates the boundary between the regions of $N_{min}=1$ and 2.}
\label{fig:5}
\end{figure}

Furthermore, we use the \citeA{Young2002} model to calculate the pitch angle diffusion coefficients ($D_{\alpha\alpha}$) driven by FLC scattering, as shown in Figure \ref{fig:5}. The format of each panel is the same as Figure \ref{fig:4}. We choose the equatorial pitch angle of the proton as 60$^{\circ}$. Even if this value of pitch angle may not be sufficient to form a bifurcated orbit for particles in certain dayside off-equatorial regions, i.e., trapping them separately in the northern and southern hemispheres, we can still obtain the influences of FLC scattering on them at the magnetic minima in both hemispheres. We present the influences in a manner separated between the northern and southern hemispheres. We speculate that if both magnetic minima in the southern and northern hemispheres are sufficient to induce significant FLC scattering effects on particles, their cumulative impact may be obtained through summation, which awaits further analysis. 

The pitch angle diffusion coefficients due to FLC scattering are highly correlated with the $Kp$ index. As the $Kp$ index increases, the effects of FLC scattering become stronger. Moreover, the asymmetry of the north-south hemisphere on off-equatorial minima regions of $D_{\alpha\alpha}$ is similar to that of $\varepsilon$. At $Kp=1$ and 3, the magnitudes of $D_{\alpha\alpha}$ are about $10^{-6}$ to $10^{-5}\,s^{-1}$ near the dusk and dawn sector, however, the magnitude of $D_{\alpha\alpha}$ can reach the $10^{-3}\,s^{-1}$ at the noon side. At $Kp=6$, the region of $D_{\alpha\alpha} \sim 10^{-3}\,s^{-1}$ near dusk and dawn sector is more expanded and is connected to the noon side for the southern off-equatorial minima regions. Though the region of dual $B$ minimum is narrow, the diffusion coefficients at noon are stronger than those at dusk and dawn sectors.

\subsection{Dipole tilt angle effects}

In the above calculation, we utilize -20$^\circ$
as the dipole tilt angle of the T89 field. The dipole tilt angle reflects the inclination of the magnetic axis with respect to the Sun. Both the sign and magnitude of the dipole tilt angle significantly influence the configuration of the dayside magnetosphere. Therefore, it is necessary to determine the probability distribution of dipole tilt angle. This information can reveal the most common magnetospheric configurations associated with specific dipole tilt angle values. Here, we statistically analyze the dipole tilt angle for each hour of every day from January 2002 to December 2012 in the T89 field, 11 years about a solar cycle. 

\begin{figure}
\centering
\noindent\includegraphics[width=0.6\textwidth]{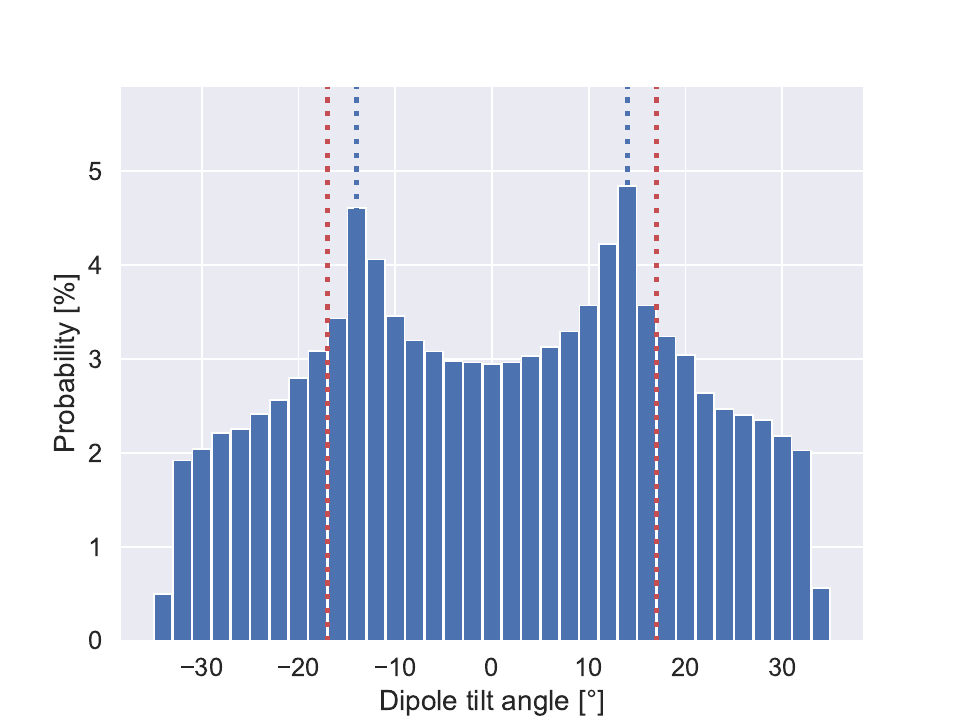}
\caption{The probability distribution of dipole tilt angles is displayed, with the red dashed line representing the median of both positive and negative dipole tilt angles, and the blue dashed line indicating the dipole tilt angle values corresponding to the bimodal peaks.}
\label{fig:6}
\end{figure}

In Figure \ref{fig:6}, we present the distribution of dipole tilt angle. Given that the angle between the magnetic axis with the rotation axis is 10.5$^\circ$, and the angle between the geographic equator with the ecliptic plane is 23.5$^\circ$, the range of dipole tilt angle is thus from -34$^\circ$ to 34$^\circ$. We mark the median values of positive and negative dipole tilt angles with blue dashed lines, while the red dashed line represents the dipole tilt angle corresponding to the peak probability. We can see that, except for dipole tilt angle values equal to about $\pm 34^\circ$, the probability corresponding to each dipole tilt angle is relatively evenly distributed, with the peak probability only about twice as high as the other probability. With an understanding of the distribution of dipole tilt angle, we can proceed to study quantitatively the effects of FLC scattering under different magnetic field configurations corresponding to various dipole tilt angle values.

\begin{figure}
\centering
\noindent\includegraphics[width=\textwidth]{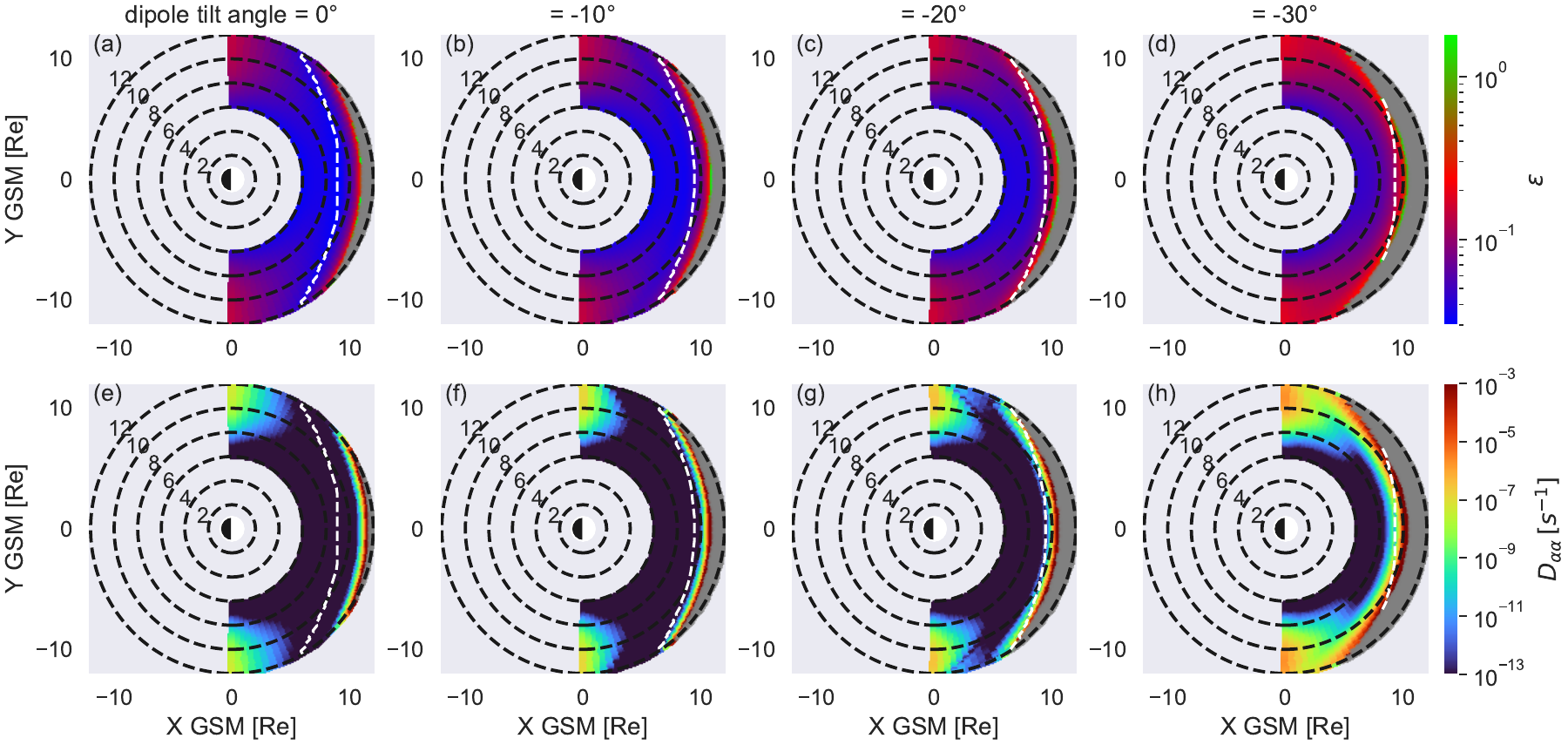}
\caption{Distribution of adiabatic parameter and pitch angle diffusion coefficients due to FLC scattering at different dipole tilt angles under $Kp=1$. The white dashed line indicates the boundary between the regions of $N_{min}=1$ and 2.}
\label{fig:7}
\end{figure}

Figure \ref{fig:7} demonstrates the distribution of $\varepsilon$ (the first row) and the pitch angle diffusion coefficients $D_{\alpha\alpha}$ (the second row) at different dipole tilt angle $0, \,-10,\,-20,\,-30^{\circ}$, respectively. The $Kp$ index is set to be 1. The impact of FLC scattering on the southern off-equatorial minimum region is much weaker than (except for cases where the dipole tilt angle is $0^{\circ}$, indicating north-south symmetry). Consequently, the results for the off-equatorial minimum regions are based on the northern off-equatorial $B$ minimum. The white dashed lines are added to show the boundary between $N_{min} =1 $ and 2. As the absolute value of the dipole tilt angle increases, the regions of off-equatorial minima (outside the white dashed line) become narrow. However, both the magnitude of $\varepsilon$ and $D_{\alpha\alpha}$ increase across all MLT sectors, including not only the dusk or dawn sector but also the noon sector. Significantly, the magnitude of $D_{\alpha\alpha}$ on the noon sector of Figure \ref{fig:7}(h) can reach $10^{-3}\,s^{-1}$. Therefore, the dipole tilt angle plays an important role in determining the extent and magnitude of FLC scattering in the dayside magnetosphere. Considering the fairly equal probability of getting different dipole tilt angles, the influences of the inclination of the magnetic axis with respect to the Sun on FLC scattering cannot be ignored.


\section{Summary and Discussion}
\label{discuss}
In this work, we investigate the impacts of FLC scattering on energetic protons in the dayside magnetosphere. By tracing the field lines of the T89 model, the off-equatorial magnetic minima regions, characterized as a W-shaped field configuration, can be determined. Considering the particles may undergo drift orbit bifurcation in the dayside off-equatorial minima regions, separate investigations of FLC scattering are performed for the northern and southern hemispheres. To quantify the proton diffusion rate by FLC scattering, we use the \citeA{Young2002} empirical model to calculate the pitch angle diffusion coefficients. In addition to the geomagnetic activity indicated by the $Kp$ index, the dipole tilt angle modulates the magnetic field configuration of the dayside magnetosphere. We demonstrate that the dipole tilt angle influences the intensity and range of FLC scattering in off-equatorial minima regions. The main conclusions can be summarized as follows:
\begin{enumerate}

  \item In the dayside off-equatorial minima regions, where radial distance reaches $7\,-\,10\,R_E$ at noon, we demonstrate that the ring current protons can be diffused by FLC pitch angle scattering. Under the configurations of the T89 model, the radial extent of FLC scattering on the noon side is less than $1\,R_E$. The intensity of FLC scattering is stronger at farther distances, which is closely related to the radial distribution of the adiabatic parameter.

  \item  By calculating the adiabatic parameter, bounce period, and applying the \citeA{Young2002} empirical model, the pitch angle diffusion coefficients driven by FLC scattering of the proton are obtained. The effects of FLC scattering are significant in dayside off-equatorial magnetic minima regions for ring current protons, in which the diffusion coefficients reach magnitudes on the order of $10^{-3} s^{-1}$ where L$ \sim 9.5\,R_\text{E}$.  

  \item The diffusion rates driven by FLC scattering in off-equatorial minima regions show remarkable hemisphere differences. Although the strengths of $B$ between dual magnetic minima are the same order of magnitude, the pitch angle diffusion coefficients differ by several orders of magnitude. This result indicates the significance of magnetic field configuration in the process of dayside FLC scattering. 

  \item On the $Z=0$ plane of GSM coordinates, the distribution of adiabatic parameters and diffusion coefficients due to FLC scattering are calculated. These results come not only from the regions of the U-shaped field lines but also from regions of off-equatorial magnetic minima. With the $Kp$ index increase, the off-equatorial minima regions contract inward. Besides, the magnitude of pitch angle diffusion coefficients due to FLC scattering becomes larger. During geomagnetic active time, the effects of FLC scattering are pronounced on the dusk or dawn side of U-shaped regions and the noon side of off-equatorial minima regions, with the noon side exhibiting hemispheric asymmetry. 

  \item The hemispheric asymmetry of FLC scattering in off-equatorial minima regions arises from the inclination of the magnetic axis with respect to the Sun. At the same $Kp$ condition, the more inclined the magnetic axis (the larger the absolute value of dipole tilt angle), the stronger and broader the FLC scattering in dayside off-equatorial regions. Additionally, for dipole tilt angles between -32$^\circ$ and 32$^\circ$, the probability associated with each tilt angle is relatively evenly distributed. These results indicate that hemispheric asymmetry in FLC scattering is a common occurrence.


\end{enumerate}

It should be pointed out that the \citeA{Young2002} model is appropriate for situations where particles are only scattered once in a half bounce period, such as the dipolar inner magnetosphere and a simple current sheet in the magnetotail \cite{Shabansky1972,Delcourt1996,Young2008}. The diffusion coefficient calculation is based on this assumption. However, in the off-equatorial minima regions, the particles may be scattered twice in a half bounce period due to the presence of dual magnetic minima. To avoid the difficulties of \citeA{Young2002} model, in this paper, separate investigations of FLC scattering are performed for the northern and southern hemispheres. Namely, we consider the process of FLC scattering occurring in different hemispheres, and we do not investigate the net effects of FLC scattering by combining two hemispheres. This will be left for future studies.

\section*{Data Availability Statement}
The simulation data produced by this work can be found at https://doi.org/10.5281/zenodo.10662712. 

\acknowledgments
This work is supported by Natural Science Foundation of Shandong Province (ZR2023MD078) and National Natural Science Foundation of China (42374194).




\bibliography{off-equator}

\begin{thebibliography}{}

\bibitem [\protect \citeauthoryear {%
Anderson%
, Decker%
, Paschalidis%
\BCBL {}\ \BBA {} Sarris%
}{%
Anderson%
\ \protect \BOthers {.}}{%
{\protect \APACyear {1997}}%
}]{%
Anderson1997}
\APACinsertmetastar {%
Anderson1997}%
\begin{APACrefauthors}%
Anderson, B\BPBI J.%
, Decker, R\BPBI B.%
, Paschalidis, N\BPBI P.%
\BCBL {}\ \BBA {} Sarris, T.%
\end{APACrefauthors}%
\unskip\
\newblock
\APACrefYearMonthDay{1997}{}{}.
\newblock
{\BBOQ}\APACrefatitle {Onset of nonadiabatic particle motion in the near-Earth
  magnetotail} {Onset of nonadiabatic particle motion in the near-earth
  magnetotail}.{\BBCQ}
\newblock
\APACjournalVolNumPages{Journal of Geophysical Research: Space
  Physics}{102}{A8}{17553-17569}.
\newblock
\begin{APACrefDOI} \doi{doi.org/10.1029/97JA00798} \end{APACrefDOI}
\PrintBackRefs{\CurrentBib}

\bibitem [\protect \citeauthoryear {%
Artemyev%
\ \protect \BOthers {.}}{%
Artemyev%
\ \protect \BOthers {.}}{%
{\protect \APACyear {2013}}%
}]{%
Artemyev2013}
\APACinsertmetastar {%
Artemyev2013}%
\begin{APACrefauthors}%
Artemyev%
, A., V.%
, Orlova%
, K., G.%
, Mourenas%
, D.%
\BDBL {}V., V.%
\end{APACrefauthors}%
\unskip\
\newblock
\APACrefYearMonthDay{2013}{}{}.
\newblock
{\BBOQ}\APACrefatitle {Electron pitch-angle diffusion: resonant scattering by
  waves vs. nonadiabatic effects.} {Electron pitch-angle diffusion: resonant
  scattering by waves vs. nonadiabatic effects.}{\BBCQ}
\newblock
\APACjournalVolNumPages{Annales Geophysicae (09927689)}{}{}{}.
\newblock
\begin{APACrefDOI} \doi{10.5194/angeo-31-1485-2013} \end{APACrefDOI}
\PrintBackRefs{\CurrentBib}

\bibitem [\protect \citeauthoryear {%
Birmingham%
}{%
Birmingham%
}{%
{\protect \APACyear {1984}}%
}]{%
Birmingham1984}
\APACinsertmetastar {%
Birmingham1984}%
\begin{APACrefauthors}%
Birmingham, T\BPBI J.%
\end{APACrefauthors}%
\unskip\
\newblock
\APACrefYearMonthDay{1984}{}{}.
\newblock
{\BBOQ}\APACrefatitle {Pitch angle diffusion in the Jovian magnetodisc} {Pitch
  angle diffusion in the jovian magnetodisc}.{\BBCQ}
\newblock
\APACjournalVolNumPages{Journal of Geophysical Research: Space
  Physics}{89}{A5}{2699-2707}.
\newblock
\begin{APACrefDOI} \doi{doi.org/10.1029/JA089iA05p02699} \end{APACrefDOI}
\PrintBackRefs{\CurrentBib}

\bibitem [\protect \citeauthoryear {%
Chirikov%
}{%
Chirikov%
}{%
{\protect \APACyear {1987}}%
}]{%
Chirikov1987}
\APACinsertmetastar {%
Chirikov1987}%
\begin{APACrefauthors}%
Chirikov, B.%
\end{APACrefauthors}%
\unskip\
\newblock
\APACrefYearMonthDay{1987}{}{}.
\newblock
{\BBOQ}\APACrefatitle {Particle Dynamics in Magnetic Traps} {Particle dynamics
  in magnetic traps}.{\BBCQ}
\newblock
\BIn{} \APACrefbtitle {Reviews of Plasma Physics} {Reviews of plasma physics}\
  (\BVOL~13, \BPG~1-99).
\newblock
\APACaddressPublisher{}{Consult. Bur., New York}.
\PrintBackRefs{\CurrentBib}

\bibitem [\protect \citeauthoryear {%
Delcourt%
, Martin~Jr.%
\BCBL {}\ \BBA {} Alem%
}{%
Delcourt%
\ \protect \BOthers {.}}{%
{\protect \APACyear {1994}}%
}]{%
Delcourt1994}
\APACinsertmetastar {%
Delcourt1994}%
\begin{APACrefauthors}%
Delcourt, D\BPBI C.%
, Martin~Jr., R\BPBI F.%
\BCBL {}\ \BBA {} Alem, F.%
\end{APACrefauthors}%
\unskip\
\newblock
\APACrefYearMonthDay{1994}{}{}.
\newblock
{\BBOQ}\APACrefatitle {A simple model of magnetic moment scattering in a field
  reversal} {A simple model of magnetic moment scattering in a field
  reversal}.{\BBCQ}
\newblock
\APACjournalVolNumPages{Geophysical Research Letters}{21}{14}{1543-1546}.
\newblock
\begin{APACrefDOI} \doi{doi.org/10.1029/94GL01291} \end{APACrefDOI}
\PrintBackRefs{\CurrentBib}

\bibitem [\protect \citeauthoryear {%
Delcourt%
, Sauvaud%
, Martin~Jr.%
\BCBL {}\ \BBA {} Moore%
}{%
Delcourt%
\ \protect \BOthers {.}}{%
{\protect \APACyear {1996}}%
}]{%
Delcourt1996}
\APACinsertmetastar {%
Delcourt1996}%
\begin{APACrefauthors}%
Delcourt, D\BPBI C.%
, Sauvaud, J\BHBI A.%
, Martin~Jr., R\BPBI F.%
\BCBL {}\ \BBA {} Moore, T\BPBI E.%
\end{APACrefauthors}%
\unskip\
\newblock
\APACrefYearMonthDay{1996}{}{}.
\newblock
{\BBOQ}\APACrefatitle {On the nonadiabatic precipitation of ions from the
  near-Earth plasma sheet} {On the nonadiabatic precipitation of ions from the
  near-earth plasma sheet}.{\BBCQ}
\newblock
\APACjournalVolNumPages{Journal of Geophysical Research: Space
  Physics}{101}{A8}{17409-17418}.
\newblock
\begin{APACrefDOI} \doi{doi.org/10.1029/96JA01006} \end{APACrefDOI}
\PrintBackRefs{\CurrentBib}

\bibitem [\protect \citeauthoryear {%
Dubyagin%
, Ganushkina%
\BCBL {}\ \BBA {} Sergeev%
}{%
Dubyagin%
\ \protect \BOthers {.}}{%
{\protect \APACyear {2018}}%
}]{%
Dubyagin2018}
\APACinsertmetastar {%
Dubyagin2018}%
\begin{APACrefauthors}%
Dubyagin, S.%
, Ganushkina, N\BPBI Y.%
\BCBL {}\ \BBA {} Sergeev, V.%
\end{APACrefauthors}%
\unskip\
\newblock
\APACrefYearMonthDay{2018}{}{}.
\newblock
{\BBOQ}\APACrefatitle {Formation of 30 ke{V} Proton Isotropic Boundaries During
  Geomagnetic Storms} {Formation of 30 ke{V} proton isotropic boundaries during
  geomagnetic storms}.{\BBCQ}
\newblock
\APACjournalVolNumPages{Journal of Geophysical Research: Space
  Physics}{123}{5}{3436-3459}.
\newblock
\begin{APACrefDOI} \doi{https://doi.org/10.1002/2017JA024587} \end{APACrefDOI}
\PrintBackRefs{\CurrentBib}

\bibitem [\protect \citeauthoryear {%
Gilson%
, Raeder%
, Donovan%
, Ge%
\BCBL {}\ \BBA {} Kepko%
}{%
Gilson%
\ \protect \BOthers {.}}{%
{\protect \APACyear {2012}}%
}]{%
Gilson2012}
\APACinsertmetastar {%
Gilson2012}%
\begin{APACrefauthors}%
Gilson, M\BPBI L.%
, Raeder, J.%
, Donovan, E.%
, Ge, Y\BPBI S.%
\BCBL {}\ \BBA {} Kepko, L.%
\end{APACrefauthors}%
\unskip\
\newblock
\APACrefYearMonthDay{2012}{}{}.
\newblock
{\BBOQ}\APACrefatitle {Global simulation of proton precipitation due to field
  line curvature during substorms} {Global simulation of proton precipitation
  due to field line curvature during substorms}.{\BBCQ}
\newblock
\APACjournalVolNumPages{Journal of Geophysical Research: Space
  Physics}{117}{A5}{}.
\newblock
\begin{APACrefDOI} \doi{https://doi.org/10.1029/2012JA017562} \end{APACrefDOI}
\PrintBackRefs{\CurrentBib}

\bibitem [\protect \citeauthoryear {%
Gray%
\ \BBA {} Lee%
}{%
Gray%
\ \BBA {} Lee%
}{%
{\protect \APACyear {1982}}%
}]{%
Gray1982}
\APACinsertmetastar {%
Gray1982}%
\begin{APACrefauthors}%
Gray, P\BPBI C.%
\BCBT {}\ \BBA {} Lee, L\BPBI C.%
\end{APACrefauthors}%
\unskip\
\newblock
\APACrefYearMonthDay{1982}{}{}.
\newblock
{\BBOQ}\APACrefatitle {Particle pitch angle diffusion due to nonadiabatic
  effects in the plasma sheet} {Particle pitch angle diffusion due to
  nonadiabatic effects in the plasma sheet}.{\BBCQ}
\newblock
\APACjournalVolNumPages{Journal of Geophysical Research: Space
  Physics}{87}{A9}{7445-7452}.
\newblock
\begin{APACrefDOI} \doi{doi.org/10.1029/JA087iA09p07445} \end{APACrefDOI}
\PrintBackRefs{\CurrentBib}

\bibitem [\protect \citeauthoryear {%
Hastie%
, Taylor%
\BCBL {}\ \BBA {} Haas%
}{%
Hastie%
\ \protect \BOthers {.}}{%
{\protect \APACyear {1967}}%
}]{%
Hastie1967}
\APACinsertmetastar {%
Hastie1967}%
\begin{APACrefauthors}%
Hastie, R\BPBI J.%
, Taylor, J\BPBI B.%
\BCBL {}\ \BBA {} Haas, F\BPBI A.%
\end{APACrefauthors}%
\unskip\
\newblock
\APACrefYearMonthDay{1967}{}{}.
\newblock
{\BBOQ}\APACrefatitle {Adiabatic invariants and the equilibrium of magnetically
  trapped particles} {Adiabatic invariants and the equilibrium of magnetically
  trapped particles}.{\BBCQ}
\newblock
\APACjournalVolNumPages{Annals of Physics}{41}{2}{302-338}.
\newblock
\begin{APACrefDOI} \doi{doi.org/10.1016/0003-4916(67)90237-0} \end{APACrefDOI}
\PrintBackRefs{\CurrentBib}

\bibitem [\protect \citeauthoryear {%
Huang%
, Tu%
\BCBL {}\ \BBA {} Eshetu%
}{%
Huang%
\ \protect \BOthers {.}}{%
{\protect \APACyear {2022}}%
}]{%
Huang2022}
\APACinsertmetastar {%
Huang2022}%
\begin{APACrefauthors}%
Huang, J.%
, Tu, W.%
\BCBL {}\ \BBA {} Eshetu, W\BPBI W.%
\end{APACrefauthors}%
\unskip\
\newblock
\APACrefYearMonthDay{2022}{}{}.
\newblock
{\BBOQ}\APACrefatitle {Modeling the Effects of Drift Orbit Bifurcation on
  Radiation Belt Electrons} {Modeling the effects of drift orbit bifurcation on
  radiation belt electrons}.{\BBCQ}
\newblock
\APACjournalVolNumPages{Journal of Geophysical Research: Space
  Physics}{127}{11}{e2022JA030827}.
\newblock
\begin{APACrefURL}
  \url{https://agupubs.onlinelibrary.wiley.com/doi/abs/10.1029/2022JA030827}
  \end{APACrefURL}
\newblock
\APACrefnote{e2022JA030827 2022JA030827}
\newblock
\begin{APACrefDOI} \doi{https://doi.org/10.1029/2022JA030827} \end{APACrefDOI}
\PrintBackRefs{\CurrentBib}

\bibitem [\protect \citeauthoryear {%
Laundal%
\ \BBA {} Richmond%
}{%
Laundal%
\ \BBA {} Richmond%
}{%
{\protect \APACyear {2017}}%
}]{%
laundal2017}
\APACinsertmetastar {%
laundal2017}%
\begin{APACrefauthors}%
Laundal, K\BPBI M.%
\BCBT {}\ \BBA {} Richmond, A\BPBI D.%
\end{APACrefauthors}%
\unskip\
\newblock
\APACrefYearMonthDay{2017}{{\APACmonth{03}}}{}.
\newblock
{\BBOQ}\APACrefatitle {Magnetic {Coordinate} {Systems}} {Magnetic {Coordinate}
  {Systems}}.{\BBCQ}
\newblock
\APACjournalVolNumPages{Space Science Reviews}{206}{1}{27--59}.
\newblock
\begin{APACrefURL}
  [{2023-07-26}]\url{https://link.springer.com/article/10.1007/s11214-016-0275-y}
  \end{APACrefURL}
\newblock
\APACrefnote{Company: Springer Distributor: Springer Institution: Springer
  Label: Springer Number: 1 Publisher: Springer Netherlands}
\newblock
\begin{APACrefDOI} \doi{10.1007/s11214-016-0275-y} \end{APACrefDOI}
\PrintBackRefs{\CurrentBib}

\bibitem [\protect \citeauthoryear {%
{Lyons}%
\ \BBA {} {Williams}%
}{%
{Lyons}%
\ \BBA {} {Williams}%
}{%
{\protect \APACyear {1984}}%
}]{%
Lyons1984}
\APACinsertmetastar {%
Lyons1984}%
\begin{APACrefauthors}%
{Lyons}, L\BPBI R.%
\BCBT {}\ \BBA {} {Williams}, D\BPBI J.%
\end{APACrefauthors}%
\unskip\
\newblock
\APACrefYear{1984}.
\newblock
\APACrefbtitle {Quantitative aspects of magnetospheric physics} {Quantitative
  aspects of magnetospheric physics}.
\newblock
\APACaddressPublisher{New York}{Springer}.
\PrintBackRefs{\CurrentBib}

\bibitem [\protect \citeauthoryear {%
Ma%
, Yu%
, Tian%
\BCBL {}\ \BBA {} Cao%
}{%
Ma%
\ \protect \BOthers {.}}{%
{\protect \APACyear {2022}}%
}]{%
Ma2022}
\APACinsertmetastar {%
Ma2022}%
\begin{APACrefauthors}%
Ma, L.%
, Yu, Y.%
, Tian, X.%
\BCBL {}\ \BBA {} Cao, J.%
\end{APACrefauthors}%
\unskip\
\newblock
\APACrefYearMonthDay{2022}{}{}.
\newblock
{\BBOQ}\APACrefatitle {An Empirical Model of the Proton Isotropic Boundary
  (IB)} {An empirical model of the proton isotropic boundary (ib)}.{\BBCQ}
\newblock
\APACjournalVolNumPages{Journal of Geophysical Research: Space
  Physics}{127}{9}{e2022JA030843}.
\newblock
\begin{APACrefDOI} \doi{https://doi.org/10.1029/2022JA030843} \end{APACrefDOI}
\PrintBackRefs{\CurrentBib}

\bibitem [\protect \citeauthoryear {%
Northrop%
}{%
Northrop%
}{%
{\protect \APACyear {1964}}%
}]{%
Northrop1964}
\APACinsertmetastar {%
Northrop1964}%
\begin{APACrefauthors}%
Northrop, T\BPBI G.%
\end{APACrefauthors}%
\unskip\
\newblock
\APACrefYearMonthDay{1964}{}{}.
\newblock
{\BBOQ}\APACrefatitle {The Adiabatic Motion of Charged Particles} {The
  adiabatic motion of charged particles}.{\BBCQ}
\newblock
\APACjournalVolNumPages{American Journal of Physics}{32}{10}{807-807}.
\newblock
\begin{APACrefDOI} \doi{10.1119/1.1969867} \end{APACrefDOI}
\PrintBackRefs{\CurrentBib}

\bibitem [\protect \citeauthoryear {%
Ozturk%
\ \BBA {} Wolf%
}{%
Ozturk%
\ \BBA {} Wolf%
}{%
{\protect \APACyear {2007}}%
}]{%
Ozturk2007}
\APACinsertmetastar {%
Ozturk2007}%
\begin{APACrefauthors}%
Ozturk, M\BPBI K.%
\BCBT {}\ \BBA {} Wolf, R\BPBI A.%
\end{APACrefauthors}%
\unskip\
\newblock
\APACrefYearMonthDay{2007}{}{}.
\newblock
{\BBOQ}\APACrefatitle {Bifurcation of drift shells near the dayside
  magnetopause} {Bifurcation of drift shells near the dayside
  magnetopause}.{\BBCQ}
\newblock
\APACjournalVolNumPages{Journal of Geophysical Research: Space
  Physics}{112}{A7}{}.
\newblock
\begin{APACrefURL}
  \url{https://agupubs.onlinelibrary.wiley.com/doi/abs/10.1029/2006JA012102}
  \end{APACrefURL}
\newblock
\begin{APACrefDOI} \doi{https://doi.org/10.1029/2006JA012102} \end{APACrefDOI}
\PrintBackRefs{\CurrentBib}

\bibitem [\protect \citeauthoryear {%
Roederer%
}{%
Roederer%
}{%
{\protect \APACyear {1970}}%
}]{%
Roederer1970}
\APACinsertmetastar {%
Roederer1970}%
\begin{APACrefauthors}%
Roederer, J\BPBI G.%
\end{APACrefauthors}%
\unskip\
\newblock
\APACrefYear{1970}.
\newblock
\APACrefbtitle {Dynamics of Geomagnetically Trapped Radiation} {Dynamics of
  geomagnetically trapped radiation}.
\newblock
\APACaddressPublisher{New York}{Springer}.
\PrintBackRefs{\CurrentBib}

\bibitem [\protect \citeauthoryear {%
Shabansky%
}{%
Shabansky%
}{%
{\protect \APACyear {1972}}%
}]{%
Shabansky1972}
\APACinsertmetastar {%
Shabansky1972}%
\begin{APACrefauthors}%
Shabansky, V\BPBI P.%
\end{APACrefauthors}%
\unskip\
\newblock
\APACrefYearMonthDay{1972}{}{}.
\newblock
{\BBOQ}\APACrefatitle {Phenomenon in the near-Earth space (p.3)(in Russian)}
  {Phenomenon in the near-earth space (p.3)(in russian)}.{\BBCQ}
\newblock

\PrintBackRefs{\CurrentBib}

\bibitem [\protect \citeauthoryear {%
Shabansky%
\ \BBA {} Antonova%
}{%
Shabansky%
\ \BBA {} Antonova%
}{%
{\protect \APACyear {1968}}%
}]{%
Shabansky1968}
\APACinsertmetastar {%
Shabansky1968}%
\begin{APACrefauthors}%
Shabansky, V\BPBI P.%
\BCBT {}\ \BBA {} Antonova, A\BPBI E.%
\end{APACrefauthors}%
\unskip\
\newblock
\APACrefYearMonthDay{1968}{}{}.
\newblock
{\BBOQ}\APACrefatitle {Topology of particle drift shells in the Earth's
  magnetosphere (English translation)} {Topology of particle drift shells in
  the earth's magnetosphere (english translation)}.{\BBCQ}
\newblock
\APACjournalVolNumPages{Geomagnetism and Aeronomy}{31(3)}{}{536–539}.
\PrintBackRefs{\CurrentBib}

\bibitem [\protect \citeauthoryear {%
Tsyganenko%
}{%
Tsyganenko%
}{%
{\protect \APACyear {1989}}%
}]{%
Tsyganenko1989}
\APACinsertmetastar {%
Tsyganenko1989}%
\begin{APACrefauthors}%
Tsyganenko, N.%
\end{APACrefauthors}%
\unskip\
\newblock
\APACrefYearMonthDay{1989}{}{}.
\newblock
{\BBOQ}\APACrefatitle {A magnetospheric magnetic field model with a warped tail
  current sheet} {A magnetospheric magnetic field model with a warped tail
  current sheet}.{\BBCQ}
\newblock
\APACjournalVolNumPages{Planetary and Space Science}{37}{1}{5-20}.
\newblock
\begin{APACrefDOI} \doi{doi.org/10.1016/0032-0633(89)90066-4} \end{APACrefDOI}
\PrintBackRefs{\CurrentBib}

\bibitem [\protect \citeauthoryear {%
Young%
, Denton%
, Anderson%
\BCBL {}\ \BBA {} Hudson%
}{%
Young%
\ \protect \BOthers {.}}{%
{\protect \APACyear {2002}}%
}]{%
Young2002}
\APACinsertmetastar {%
Young2002}%
\begin{APACrefauthors}%
Young, S\BPBI L.%
, Denton, R\BPBI E.%
, Anderson, B\BPBI J.%
\BCBL {}\ \BBA {} Hudson, M\BPBI K.%
\end{APACrefauthors}%
\unskip\
\newblock
\APACrefYearMonthDay{2002}{}{}.
\newblock
{\BBOQ}\APACrefatitle {Empirical model for μ scattering caused by field line
  curvature in a realistic magnetosphere} {Empirical model for μ scattering
  caused by field line curvature in a realistic magnetosphere}.{\BBCQ}
\newblock
\APACjournalVolNumPages{Journal of Geophysical Research: Space
  Physics}{107}{A6}{SMP 3-1-SMP 3-9}.
\newblock
\begin{APACrefDOI} \doi{doi.org/10.1029/2000JA000294} \end{APACrefDOI}
\PrintBackRefs{\CurrentBib}

\bibitem [\protect \citeauthoryear {%
Young%
, Denton%
, Anderson%
\BCBL {}\ \BBA {} Hudson%
}{%
Young%
\ \protect \BOthers {.}}{%
{\protect \APACyear {2008}}%
}]{%
Young2008}
\APACinsertmetastar {%
Young2008}%
\begin{APACrefauthors}%
Young, S\BPBI L.%
, Denton, R\BPBI E.%
, Anderson, B\BPBI J.%
\BCBL {}\ \BBA {} Hudson, M\BPBI K.%
\end{APACrefauthors}%
\unskip\
\newblock
\APACrefYearMonthDay{2008}{}{}.
\newblock
{\BBOQ}\APACrefatitle {Magnetic field line curvature induced pitch angle
  diffusion in the inner magnetosphere} {Magnetic field line curvature induced
  pitch angle diffusion in the inner magnetosphere}.{\BBCQ}
\newblock
\APACjournalVolNumPages{Journal of Geophysical Research: Space
  Physics}{113}{A3}{}.
\newblock
\begin{APACrefDOI} \doi{doi.org/10.1029/2006JA012133} \end{APACrefDOI}
\PrintBackRefs{\CurrentBib}

\bibitem [\protect \citeauthoryear {%
Yu%
, Tian%
\BCBL {}\ \BBA {} Jordanova%
}{%
Yu%
\ \protect \BOthers {.}}{%
{\protect \APACyear {2020}}%
}]{%
Yu2020}
\APACinsertmetastar {%
Yu2020}%
\begin{APACrefauthors}%
Yu, Y.%
, Tian, X.%
\BCBL {}\ \BBA {} Jordanova, V\BPBI K.%
\end{APACrefauthors}%
\unskip\
\newblock
\APACrefYearMonthDay{2020}{}{}.
\newblock
{\BBOQ}\APACrefatitle {The Effects of Field Line Curvature (FLC) Scattering on
  Ring Current Dynamics and Isotropic Boundary} {The effects of field line
  curvature (flc) scattering on ring current dynamics and isotropic
  boundary}.{\BBCQ}
\newblock
\APACjournalVolNumPages{Journal of Geophysical Research: Space
  Physics}{125}{8}{e2020JA027830}.
\newblock
\APACrefnote{e2020JA027830 10.1029/2020JA027830}
\newblock
\begin{APACrefDOI} \doi{https://doi.org/10.1029/2020JA027830} \end{APACrefDOI}
\PrintBackRefs{\CurrentBib}

\bibitem [\protect \citeauthoryear {%
Yue%
\ \protect \BOthers {.}}{%
Yue%
\ \protect \BOthers {.}}{%
{\protect \APACyear {2014}}%
}]{%
Yue2014}
\APACinsertmetastar {%
Yue2014}%
\begin{APACrefauthors}%
Yue, C.%
, Wang, C\BHBI P.%
, Lyons, L.%
, Liang, J.%
, Donovan, E\BPBI F.%
, Zaharia, S\BPBI G.%
\BCBL {}\ \BBA {} Henderson, M.%
\end{APACrefauthors}%
\unskip\
\newblock
\APACrefYearMonthDay{2014}{}{}.
\newblock
{\BBOQ}\APACrefatitle {Current sheet scattering and ion isotropic boundary
  under 3-D empirical force-balanced magnetic field} {Current sheet scattering
  and ion isotropic boundary under 3-d empirical force-balanced magnetic
  field}.{\BBCQ}
\newblock
\APACjournalVolNumPages{Journal of Geophysical Research: Space
  Physics}{119}{10}{8202-8211}.
\newblock
\begin{APACrefDOI} \doi{https://doi.org/10.1002/2014JA020172} \end{APACrefDOI}
\PrintBackRefs{\CurrentBib}

\bibitem [\protect \citeauthoryear {%
Zhu%
, Yu%
, Tian%
, Shreedevi%
\BCBL {}\ \BBA {} Jordanova%
}{%
Zhu%
\ \protect \BOthers {.}}{%
{\protect \APACyear {2021}}%
}]{%
Zhu2021}
\APACinsertmetastar {%
Zhu2021}%
\begin{APACrefauthors}%
Zhu, M.%
, Yu, Y.%
, Tian, X.%
, Shreedevi, P\BPBI R.%
\BCBL {}\ \BBA {} Jordanova, V\BPBI K.%
\end{APACrefauthors}%
\unskip\
\newblock
\APACrefYearMonthDay{2021}{}{}.
\newblock
{\BBOQ}\APACrefatitle {On the Ion Precipitation due to Field Line Curvature
  (FLC) and EMIC Wave Scattering and Their Subsequent Impact on Ionospheric
  Electrodynamics} {On the ion precipitation due to field line curvature (flc)
  and emic wave scattering and their subsequent impact on ionospheric
  electrodynamics}.{\BBCQ}
\newblock
\APACjournalVolNumPages{Journal of Geophysical Research: Space
  Physics}{126}{3}{e2020JA028812}.
\newblock
\begin{APACrefDOI} \doi{https://doi.org/10.1029/2020JA028812} \end{APACrefDOI}
\PrintBackRefs{\CurrentBib}

\end{thebibliography}

\newpage

\end{document}